\documentclass[reprint,aps,aip,twocolumn]{revtex4-1}                     
\usepackage{graphicx, amsmath}
\usepackage[separate-uncertainty=true]{siunitx}
\usepackage{amssymb}
\usepackage{comment}
\usepackage[sort&compress]{natbib}
\begin{document}

	\title{Aharonov-Casher Effect and the Coherent Flux Tunneling in the Hybrid Charge Quantum Interference Device}
	\author{J. W. Dunstan}
	\affiliation{Royal Holloway, University of London, Egham Hill, Egham TW20 0EX, United Kingdom}
	\author{R. Shaikhaidarov}
	\affiliation{Royal Holloway, University of London, Egham Hill, Egham TW20 0EX, United Kingdom}
	\affiliation{National Physical Laboratory, Hampton Road, Teddington, TW11 0LW, United Kingdom}
	\author{K. H. Kim}
	\affiliation{Royal Holloway, University of London, Egham Hill, Egham TW20 0EX, United Kingdom}
	\author{A. Shesterikov}
	\affiliation{Royal Holloway, University of London, Egham Hill, Egham TW20 0EX, United Kingdom}
	\affiliation{National Physical Laboratory, Hampton Road, Teddington, TW11 0LW, United Kingdom}
    \author{S. Linzen} \affiliation{Leibniz Institute of Photonic Technology, D-07702 Jena, Germany}
     \author{E. V. Il’ichev} \affiliation{Leibniz Institute of Photonic Technology, D-07702 Jena, Germany}
	\author{I. Antonov}
	\affiliation{Royal Holloway, University of London, Egham Hill, Egham TW20 0EX, United Kingdom}
	\affiliation{National Physical Laboratory, Hampton Road, Teddington, TW11 0LW, United Kingdom}
	\author{V. N. Antonov}
	\affiliation{Royal Holloway University of London, Egham, Surrey, TW20 0EX, United Kingdom}
	\author{O. V. Astafiev}
	\affiliation{Royal Holloway University of London, Egham, Surrey, TW20 0EX, United Kingdom}

\begin{abstract}
By exploiting the Aharonov-Casher effect we demonstrate a suppression of magnetic flux tunneling in a Hybrid Charge Quantum Interference Device. The main part of this device is two Josephson junctions with a small superconducting island between them. To minimize phase fluctuations across Josephson junctions, this structure is embedded in a compact super-inductive NbN loop. The Interference between the flux tunneling paths is determined by the island-induced charge, which is controlled by an external voltage. The charge sensitive operation of the device is subjected to poisoning by the quasiparticles generated in the NbN film.

\end{abstract}
\maketitle

The Aharonov-Casher (AC) effect, the interference of fluxons around a static charge, is dual to the Aharonov-Bohm (AB) effect, the interference of the charged particles encircling a magnetic flux \cite{Aharonov1984}. The AC effect in superconducting coherent circuits has been the subject of a few works so far. The effect was experimentally demonstrated in spectroscopy of the charge and bifluxon qubits \cite{Bell2016,Kalashnikov2020}, switching current of the linear array of Josephson junctions (JJ) \cite{Pop2012}, and resistance of the JJ network \cite{Elion1993}. Freedman and Averin suggested using the AC effect to completely suppress fluxon tunneling in the JJs \cite{Friedman2002}. Partially this has been achieved in bifluxon qubit \cite{Kalashnikov2020} and Charge Quantum Interference Device (CQUID) \cite{deGraaf2018}. In the latter experiment, the JJs were replaced with the nanowires.  Despite a similar appearance in the experiment, the underlying physics of nanowires and JJs is somewhat different \cite{deGraaf2018,deGraaf2020}. In addition, complete suppression of the fluxon tunneling has not been demonstrated in both experiments.

To demonstrate AC effect, one has to ensure operation of JJs/nanowires in the regime of coherent quantum phase slip (CQPS). It requires screening the CQPS centre from the environmental noise, because the phase slip energy $E_S$ is usually tiny, few $\mu$V ($\sim$30 - 50~mK). External noise engages phase fluctuations, thus further reducing $E_S$. In RF-SQUID-type devices, where phase fluctuation is also an issue, a large superconducting loop with high inductance is used \cite{Ilichev1997}. In later work, high inductance was realized by a linear array of JJs. Alternatively, CQUID used an NbN film with extreme kinetic inductance ($L_{\square}\sim$~1.6~nH) as the material of the loop \cite{deGraaf2018}. This made the device very compact and resistant to external flux noise. 

However, there are limitations of the CQUID, which has the nanowires as CQPS centres: the width of the nanowires required by the CQPS,  $\sim 10$ nm, is at the limit of nanofabrication accuracy, which is currently about $\pm$~2~nm. There is large variation in the CQPS energy $E_S$ due to its exponential sensitivity to the  width of nanowires \cite{Peltonen2013,Golubev2001,Zaikin1997}. In contrast, JJs can be fabricated with fairly good reproducibility of parameters, better than 90~$\%$ . Currently, JJs integrated with high inductance materials are an active field of research for the engineering of superconducting coherent circuits \cite{Manucharyan2009,Kalashnikov2020, Pechenezhskiy2020}. For example NbN, TiN, or $\mathrm{InO_{x}}$  wires replace the super-inductive array of JJs in the applications where a small stray capacitance is required \cite{Shaikhaidarov2022, Shaikh2024, ShaikhAPL2024}.\par
In our work we designed the hybrid CQUID (h-CQUID), where the JJs are combined with the high kinetic inductance NbN films. The h-CQUID has the advantage of reliable control of the parameters compared to CQUID (the yield of the nanowires with proper parameters after fabrication is below 0.3), compactness of the high kinetic inductance material, and low stray capacitance.

Figure 1 displays SEM images of the sample and the experimental setup. In h-CQUID the JJs are embedded in the NbN loop. The NbN films grown by atomic layer deposition \cite{Linzen_2017}. The JJs have good galvanic contact with the NbN loop.

As will be shown below, for such a structure the interference of fluxons crossing the JJs can be controlled by the gate voltage up to complete suppression of the fluxon tunnelling, thereby increasing the functionality of the h-CQUID, and confirming the theoretical prediction \cite{Friedman2002}. 
\begin{figure}[!h]
\includegraphics[width=\columnwidth]{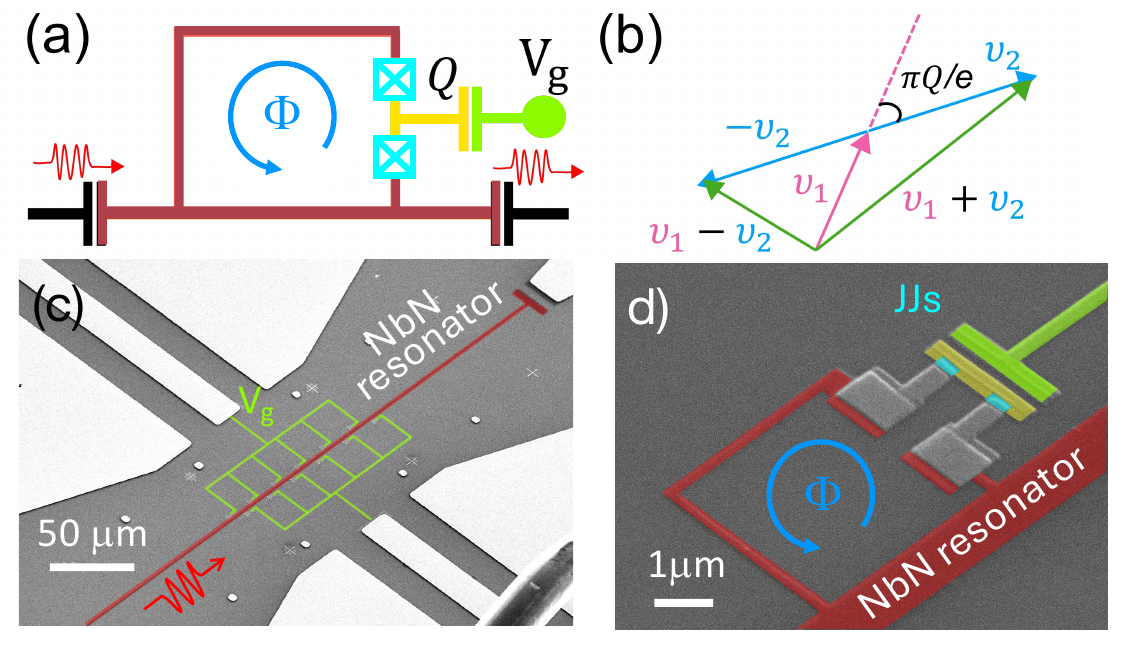}
\caption{(a) A schematic of the sample. The h-CQUID consists of the superconducting loop (venous) interrupted by two Josephson junctions (cyan). A superconducting island (yellow) between the junctions is capacitively coupled to a voltage gate. The h-CQUID shares part of the loop with $\lambda$/2 resonator. (b) A representation of addition and subtraction of phase slip phasors in Josephson junction 1 and 2. The angle between the phasors is $\pi Q/e$. With additional quasiparticle in the island, the total phase slip amplitude becomes $\lvert\nu_1 - \nu_2\rvert$. (c)  The SEM image (false colour) of multiple h-CQUIDs coupled to a $\lambda$/2 resonator (venous) and gate electrodes (green). (d) SEM image of a single h-CQUID. The picture has the same colour scheme as (a) and (c).}
\label{fig: sample}
\end{figure}

The Hamiltonian of the system in the flux basis can be written as:
\begin{equation}
    \hat{H}=-(\epsilon/2) \hat{\sigma}_z - (E_{\mathrm{S}}/2) \hat{\sigma}_x
    \label{eq: Hamiltonian1}
\end{equation} 
where $\epsilon = 2 I_{\mathrm{p}} \Delta \Phi$ is the energy difference between the two fluxon number states, $E_{\mathrm{S}}$ is the coupling energy between these states, and  $\sigma_z$ and  $\sigma_x$ are Pauli matrices. Here, $I_{\mathrm{p}}=\Phi_0/2 L_{\mathrm{k}}$ is the persistent current, $L_{\mathrm{k}}$ the kinetic inductance of the loop, $\Delta \Phi =\Phi-(N+1/2)\Phi_0$ is the deviation from the degeneracy point. The coupling energy $E_{\mathrm{S}}$, also known as the phase slip energy, is related to the tunneling rates of the fluxons, $\nu_{1,2}$, across the JJs
$E_{\mathrm{S}}(Q)=h\left|\nu_1+e^{i2\pi Q}\nu_2\right|$. In the latter equation $Q$ is the charge on a small Al island between the JJs. The charge is induced by the voltage $V_g$ at the metal gate, $Q=C_gV_g$. It changes the vector sum of the fluxon tunneling rate, see Figure 1(b) for two values of $Q$, different by the charge of electron $e$ \cite{deGraaf2018,Pop2012}. The excitation energy of the qubit is

\begin{equation}
\Delta E=\sqrt{\epsilon(\Phi)^2+E_{\mathrm{S}}(Q)^2}
\label{eq: deltaE}
\end{equation}

It depends on both the external magnetic flux $\Phi$ as in a conventional RF-SQUID, and the induced charge $Q$.

The h-CQUID is inductively coupled to a line resonator. For probing of the qubit state, we used two-tone spectroscopy with a dispersive readout at the resonant frequency of the resonator. Five samples were measured. Each of them had four h-CQUIDs coupled to the same resonator. The experimental data presented here are taken from one of the samples. Measurements were performed at \SI{12}{\milli\kelvin} in a dilution refrigerator.

\begin{figure}[htb]
\includegraphics[width=\columnwidth]{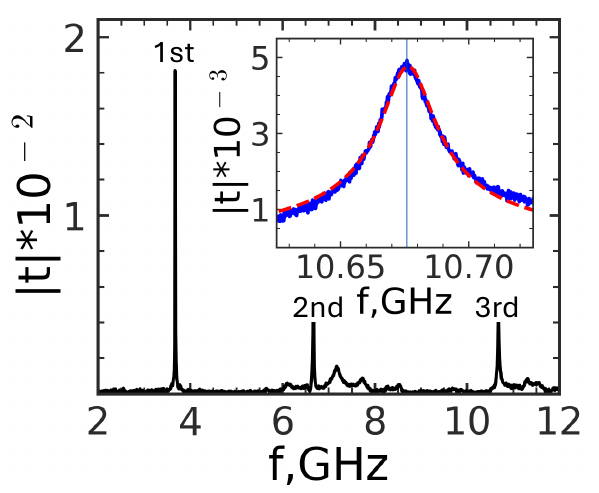}
\caption{Transmission of the NbN line resonator. There are 3 resonance modes in the spectral range of our measurement: 3.671 GHz, 6.670~GHz, 10.676 GHz. Insert: Third mode (blue) at 10.676 GHz is used for two-tone spectroscopy. Dashed red line is the Lorentzian fit. The resonance has a quality factor of $Q\approx$ 520.}
\label{fig: TransmissionMode}
\end{figure}

We probe transmission through the resonator with the Vector Network Analyser (VNA). The input microwave line has \SI{-83}{\dB} attenuation from room to low temperature. The transmitted signal is amplified by +\SI{35}{\dB} at low temperature and +\SI{35}{\dB} at room temperature before reaching the VNA input port. There are three modes of the resonator in the experimental range see Fig. \ref{fig: TransmissionMode}. Ideally, they should be equidistant with frequencies $f_{n}=nv/2L$ ($n$=1, 2, 3, $v$ is the phase velocity, and $L$ is the length of the resonator, $v$ is the phase velocity). In experiment resonance frequencies usually deviates from the calculated values, because of the qubit loops inductively coupled to the resonator. For two-tone spectroscopy we use 3$^{\text{rd}}$ resonance at \SI{10.676}{\giga\hertz}. It has a quality factor $Q\approx520$. Despite the quality factor of this resonance concedes to that of 1$^{\text{st}}$ and 2$^{\text{nd}}$ modes, it has the strongest coupling to the h-CQUID under analysis.

\begin{figure}[htb!]
\includegraphics[width=\columnwidth]{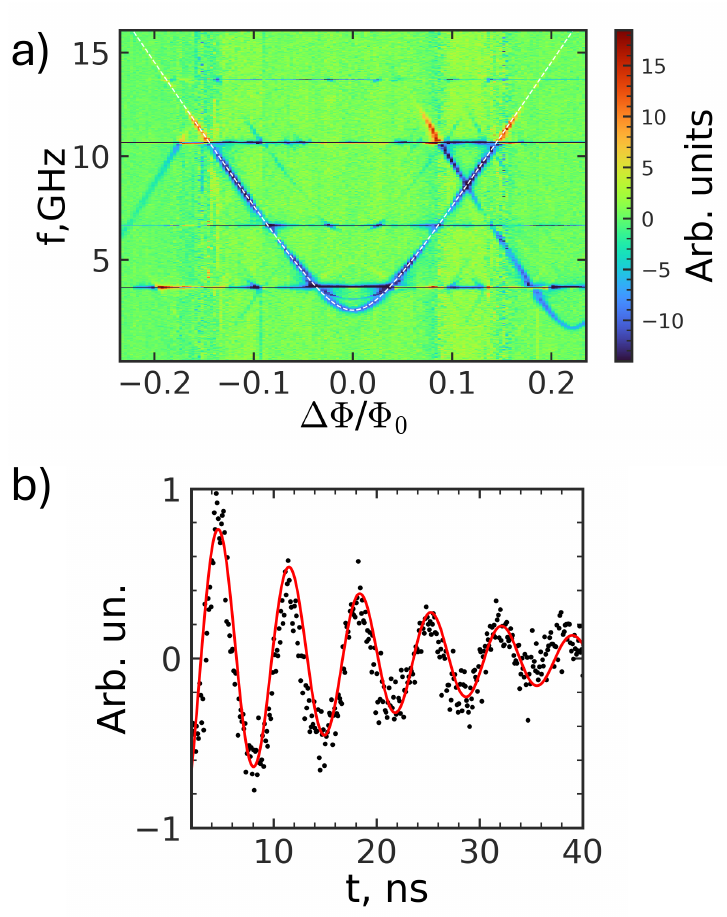}
\caption{(a) Two-tone spectroscopy of the h-CQUID at different deviation of magnetic flux $\Delta \Phi$ from the degeneracy point. The spectral line is overlaid with white dashed line of equation (\ref{eq: deltaE}). The parameters of the fit line are: $I_p$= 11.5~nA, $E_S(Q)$=~2.555~GHz. There are additional spectroscopy lines in the intensity plot (the modes of resonator, different qubits etc.) (b) Rabi oscillations of the qubit taken at the degeneracy point $\Delta \Phi$=0. The relaxation time of qubit $T_1$=20 ns. }
\label{fig: mField}
\end{figure}

Figure \ref{fig: mField}(a) shows a two-tone spectroscopy of the sample: we probe transmission of the resonator at \SI{10.676}{\giga\hertz} while sweeping the second MW signal from \SI{0.1}{GHz} to \SI{10}{GHz}. The transmission drops when the frequency of the second MW signal is equal to the excitation frequency of the qubit, $\Delta E/h$. The frequency scans at different magnetic fluxes are compiled in the intensity plot. We focus on the spectral line of the flux qubit positioned in the centre. The minimum of this spectral line at the degeneracy point $\Delta\Phi$=0 is , $f$~=~2.555~GHz. We fit the curve with (\ref{eq: deltaE} and find a persistent current of the loop $I_P\sim$~11.5~nA. It corresponds to the total inductance of the h-CQUID loop with the JJs of 90~nH. There are few additional spectroscopy lines of different nature, which are not important for our analysis: other resonances of the sample, the modes of the linear resonator, and higher excitations of the structure under study. The Rabi oscillations of the h-CQUID taken at the degeneracy point $\Delta \Phi$=0 are shown in Fig. \ref{fig: mField}(b). From the fit of the oscillations  we find the relaxation time $T_1\simeq$20~ns. It is much smaller than in the modern qubit, where $T_1$ reaches millisecond. Partially, such a low relaxation time can be attributed to a strong inductive coupling of the h-CQUID loop to the resonator.        


\begin{figure}[htb]
\includegraphics[width=\columnwidth]{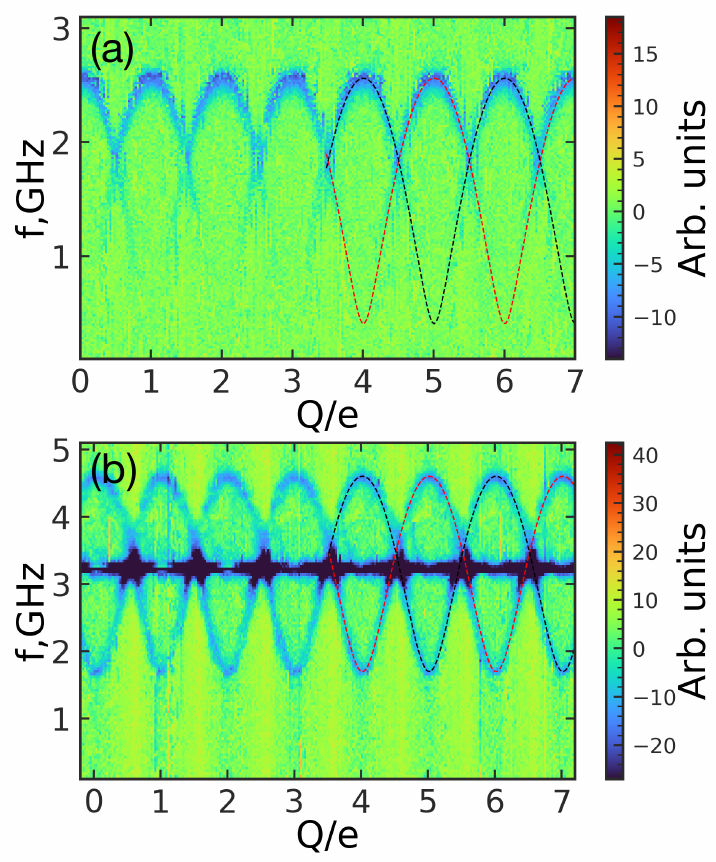}
\caption{(a) Spectroscopy of the h-CQUID under study at the degeneracy point $\Delta\Phi$=0 taken at different $Q=C_gV_g$. h-CQUID has small asymmetry of JJs, $\nu_1$=~1.48~GHz and $\nu_2$=~1.08~GHz. There are two oscillating curves with even and odd parities of charge. The period of oscillations of individual curve is $2e$. The curves are overlaid with the dashed fitting lines of Eq.~\ref{eq: deltaE}. (b) Spectroscopy of the h-CQUID with large asymmetry of JJs, $\nu_1$=~3.15~GHz and $\nu_2$=~1.45~GHz.}
\label{fig: GateSweep}
\end{figure}

There is an additional gate voltage control of the h-CQUID. We fix the magnetic flux at the degeneracy point and compile the frequency scans at different $V_g$ to the intensity plot of Fig. \ref{fig: GateSweep}(a). There are two spectral lines that oscillate with period $Q=2e$. The oscillation phase of them is exactly shifted by $e$. The effect is similar to that seen in CQUID with superconducting nanowires replacing the JJs \cite{deGraaf2018}. The oscillations are manifestations of the Aharonov-Casher effect. Near the point of destructive interference, $Q=2e(n+1/2)$, the flux tunneling rate $\nu$ should be minimal. In the ideal case of two identical JJs,  when $\nu_1=\nu_2$,  the flux tunneling should be completely suppressed, the excitation energy of the qubit should tend to zero, and the spectral lines should vanish \cite{Friedman2002}. We fit experimental curves with $\nu=|\nu_{\mathrm{1}}\pm\nu_{\mathrm{2}}e^{i 2\pi Q}|$, with $\nu_{\mathrm{1}}=\SI{1.48}{\giga\hertz}$ and $\nu_{\mathrm{2}}=\SI{1.08}{\giga\hertz}$. The minimum of $E_S/h\approx$0.4 GHz corresponds to the case when two phase slip phasors have opposite directions. The different tunneling rates are a consequence of different parameters of the JJ in h-CQUID.  We discuss this in the following section.            
The presence of two spectroscopy curves in Fig.~\ref{fig: GateSweep} is a sign of the quasi-particle poisoning of the Al island between the JJs.  We believe that these unpaired electrons are excited by the microwave noise in the highly inductive NbN loop \cite{deGraaf2018}. Quasi-particle poisoning presents a serious problem for the devices utilizing superconducting films of high kinetic inductance \cite{Shaikh2024}.     

\par 

In small JJ with $E_{\mathrm{J}}$ few times larger than $E_{\mathrm{C}}$ the tunneling rate of the fluxons can be found from 
\begin{equation}
\nu=\frac{8}{h}\sqrt{E_{\mathrm{J}}E_{\mathrm{C}}/\pi}(8E_{\mathrm{J}}/E_{\mathrm{C}})^{1/4}\exp{(-\sqrt{8E_{\mathrm{J}}/E_{\mathrm{C}}})}
\label{eq: nu}
\end{equation}
where $E_{\mathrm{C}}=e^{2}/2C$, $E_{\mathrm{J}}= R_{\mathrm{q}}\Delta/2R_{\mathrm{N}}$, and $R_{\mathrm{N}}$ are the charging energy, the Josephson energy, and the normal resistance \cite{SCHON1990,Likharev1985}. We estimate $E_{\mathrm{C}}/h\approx\SI{13.8}{\giga\hertz}$ and $E_{\mathrm{J}}/h\approx\SI{29.2}{\giga\hertz}$ using the specific capacitance $\sim5\times10^{-16} \textrm{ }\mathrm{F/\mu m^{2}}$, junction area $0.28\times0.1\textrm{ }\mathrm{}\mu m^{2}$, $\Delta_{Al}$=202~$\mu$V, and $R_{\mathrm{N}}=\SI{5.0}{\textrm{}k\Omega}$ \cite{Frank2004}. Then the fluxon tunneling rate from equation (\ref{eq: nu}) is $\nu\approx\SI{2.7}{\giga\hertz}$. It is twice larger than the experimentally found $\SI{1.48}{\giga\hertz}$ and $\SI{1.08}{\giga\hertz}$. Poor agreement can be explained by exponential sensitivities of $\nu$ to the parameters of the JJ tunnel barrier. The tunneling barrier affects the Josephson energy in a large way compared to the charging energy. The charging energy $E_C$ is determined more steadily from the geometry of the JJ. An uncertainty of less than 5~$\%$ can be expected. To be close to the experimental value, the ratio $E_J/E_C$ should be in the range of 3.0-3.2 (we fix $E_{\mathrm{C}}$ = 13.8~GHz) compared to $\sim$2.3 in our calculation. The strong sensitivity of $E_J$ to the fabrication process explains the 40~$\%$ difference between geometrically identical $\nu_1$ and $\nu_2$. To demonstrate effect of asymmetry of the JJ we show the spectroscopy curve of h-CQUID, where the tunnelling rates are quite different, $\nu_1$=~3.15~GHz and $\nu_2$=~1.45~GHz, see Fig. \ref{fig: GateSweep}(b). The overlapping areas of the JJs in this sample are only slightly different, their ratio is 2.5/2. One can clearly see the top and the bottom of the spectroscopy curves.          


\par
So far we have discussed h-CQUID using CQPS model of the JJs. It is also instructive to model h-CQUID starting from the flux qubit Hamiltonian in the basis of superconducting phase. The Hamiltonian of the superconducting loop with two identical Josephson junctions reads  
\begin{equation}H=-8E_{\mathrm{C}} \partial^2/\partial{\phi^2} -2E_{\mathrm{J}} \lvert\cos{(\phi/2)}\rvert+\frac{E_{\mathrm{L}}}{2} (\phi-\phi_{ext})^2
\label{eq: Hamiltonian2}
\end{equation}
where $E_{\mathrm{L}}=\Phi_0^2/4\pi^2 L_{\mathrm{k}}$, $\phi$ is the total superconducting phase difference of the two junctions, $\phi=\phi_{\mathrm{1}}+\phi_{\mathrm{2}}$, and $\phi_{\mathrm{ext}}=2\pi \Phi_{\mathrm{ext}}/\Phi_{\mathrm{0}}$ \cite{Shnyrkov2008}. Note that $E_{\mathrm{C}}$ and $E_{\mathrm{J}}$ are taken for a single junction, so that both the kinetic and Josephson energy terms in the Hamiltonian are approximately doubled compared to usual RF-SQUID \cite{Dmitriev2021,Peltonen2018}. Also the Hamiltonian (\ref{eq: Hamiltonian2}) assumes a well-defined phase and does not include the effect of single electron charging of the island. Therefore, we fit only the spectroscopy line in Fig. \ref{fig: mField}(b). The produce the best fit using the numerical diagonalization of the Hamiltonian (\ref{eq: Hamiltonian2}) and varying $E_J$ and $E_L$ ($E_C$ is again fixed to 13.8~GHz). The simulation gives $E_{\mathrm{J}}/h = \SI{27.9}{\giga\hertz}$, $E_{\mathrm{L}}/h=\SI{1.89}{\giga\hertz}$ ($L_k$=81.3~nH). Despite the simplicity of this model, these parameters are consistently close to the values obtained from the fit of the curves with equation (\ref{eq: deltaE}).

In conclusion, we demonstrate operation of hybrid CQUID, made of a combination of high inductive NbN loop and JJs. The yield of such h-CQUID is substantially higher than that of the CQUID made of the superconducting nanowires. The fluxons coherently tunneling through the small Josephson junctions develop interference, which is controlled with the charge induced at the island between the JJs (Aharonov-Casher effect). When the induced charge has an odd number of electron charges $e$, the interference is destructive so that the flux tunneling rate tends to zero. It is seen as a dissolution of the spectroscopy line of the h-CQUID. Although h-CQUID has better control of the fluxon tunneling rate compared to original CQUID, it does not eliminate the problem of quasiparticles poisoning of the device. The latter has its origin in the quasiparticle excitations in the NbN film under the MW signal. \par

This work was supported by EMPIR 20FUN07 SuperQuant, and UK Engineering and Physical Sciences Research Council (EPSRC) Grant No. EP/Y022637/1. S.L and E.I would like to thank the German Federal Ministry of Research, Technology and Space (BMFTR) for partial support under Grant No. 13N17121/NbNanoQ. 

\section*{Data availability}
The data generated in this study have been deposited in the Open Science
Framework repository. They can be obtained without any restriction at \href{https://osf.io/Q4dav}{https://osf.io/Q4dav}. 

\bibliography{HybridQubit}
	\bibliographystyle{unsrt}
 
\end{document}